\documentclass[aps,twocolumn,superscriptaddress,showpacs,floats]{revtex4}
\usepackage{amsmath}
\usepackage{amsfonts}
\usepackage{amssymb}
\usepackage{euscript}
\usepackage{bm}
\usepackage[dvips]{graphicx}

\begin{document}

\title{Scanning Superfluid-Turbulence Cascade by Its Low-Temperature Cutoff}

\author{Evgeny Kozik}
 \affiliation{Institute for Theoretical Physics, ETH Zurich, CH-8093 Zurich, Switzerland}
 \affiliation{Department of Physics, University of
Massachusetts, Amherst, MA 01003}
\author{Boris Svistunov}
\affiliation{Department of Physics, University of Massachusetts,
Amherst, MA 01003} \affiliation{Russian Research Center
``Kurchatov Institute'', 123182 Moscow, Russia}

\begin{abstract}
On the basis of recently proposed scenario of the transformation
of the Kolmogorov cascade into the Kelvin-wave cascade, we develop
a theory of low-temperature cutoff. The theory predicts a specific
behavior of the quantized vortex line density, $L$, controlled by
the frictional coefficient, $\alpha(T) \ll 1$, responsible for the
cutoff. The curve $\ln L(\ln \alpha)$ is found to directly reflect
the structure of the cascade, revealing four qualitatively
distinct wavenumber regions. Excellent agreement with recent
experiment by Walmsley {\it et al.} [Phys. Rev. Lett. 99, 265302
(2007)]---in which $L(T)$ has been measured down to $T \sim
0.08\,$K---implies that the scenario of low-temperature superfluid
turbulence is now experimentally validated, and allows to quantify
the Kelvin-wave cascade spectrum.

\end{abstract}


%
%
%
%
%

\maketitle

In the studies of superfluid turbulence (ST) (for introduction,
see, e.g., \cite{Donnelly, Cambridge}), the case of very low
temperatures remains intriguing and challenging
\cite{Physics_today}. A wealth of theoretical work has been
devoted to ST in this regime in the past decade and especially in
the past few years
\cite{Sv_95,Nore,Vinen2000,Kivotides,Tsubota_2002,Vinen_2003,KS_04,KS_05_vortex_phonon,Nazarenko,Lvov,KS_07},
and a detailed theoretical picture seems to emerge. In this
picture, the most interesting physics is associated with the short
length scales, where quantized nature of vortices manifests itself
and the key role is played by Kelvin waves (KWs)---the distortion
waves on quantized vortex lines. However, the feasibility of
experimental confirmation of these predictions with the current
technique seemed vague, which to many made the theory appear
purely academic. Until very recently, the only experimentally
confirmed fact had been the existence of cascades in the $T\to 0$
limit \cite{Davis, Pickett}.

In their recent remarkable work \cite{Walmsley}, Walmsley {\it et
al.} measured the vortex line density $L$ (the length of vortex
lines per unit volume), of the Kolmogorov cascade in superfluid
$^4$He down to temperatures $T \sim 0.08\,$K. They found a
striking temperature dependence of $L$ and argued that it should
be due to the Kelvin-wave cascade
\cite{Sv_95,Nore,Vinen2000,Kivotides,Tsubota_2002,Vinen_2003,KS_04,KS_05_vortex_phonon,Nazarenko,Lvov,KS_07}
being extended to higher wavenumbers with decreasing temperature.
Better understanding of this process was formulated as a challenge
for theorists.

In this Letter, we meet the above-mentioned challenge. Recently,
the two of us proposed a detailed scenario  \cite{KS_07} of how
the zero-temperature Kolmogorov cascade transforms into the pure
KW cascade. We argued that the transformation involves a chain of
three intermediate regimes and thus occupies a significant
interval in the wavenumber space.  On the basis of this
zero-temperature scenario, we now develop a theory of the
low-temperature dissipative cutoff of the cascade. We show that
the temperature dependence of $L$ comes from the dependence of the
wavelength scale $\lambda_\mathrm{cuttoff}$, at which the cascade
ceases due to the mutual friction of vortex lines with the normal
component, on the dimensionless friction coefficient $\alpha
\propto T^5~~(T\to 0)$. A very characteristic shape of the
function $\ln L(\ln \alpha)$ following from our analysis is in an
excellent agreement with the experimental results of Walmsley {\it
et al.} Remarkably, the shape of the curve $\ln L(\ln \alpha)$
directly reflects four qualitatively distinct wavenumber regions
of the cascade.
The agreement between our theory and the
data of Ref.~\cite{Walmsley} means that the scenario of
low-temperature superfluid turbulence is now experimentally
validated. Moreover, fitting the experimental data with
theoretical curves allows us to quantify the KW cascade
spectrum.

A cascade of energy---the decay mechanism of both classical and
superfluid turbulence---implies the existence of an
\textit{inertial range}---a significant range of length scales,
$\lambda_\mathrm{en} > \lambda > \lambda_\mathrm{cutoff}$, in
which all the dissipative mechanisms are weak rendering the system
essentially conservative. The long- and the short-wavelength ends
of the inertial range play a special role: almost all the energy
of the system is concentrated at $\lambda_\mathrm{en}$, while the
dissipation becomes appreciable only at $\lambda \lesssim
\lambda_\mathrm{cutoff} \equiv \lambda_\mathrm{cutoff}(T)$. The
relaxation process must lead to a transfer of the turbulent energy
towards $\lambda_\mathrm{cutoff}$, where it can be dissipated into
heat. For the cascade regime to set in, it is also necessary that
the dynamics within the inertial range is such that the
``collisional'' kinetic time $\tau_{\rm coll} (\lambda)$, i.e. the
time between elementary events of energy exchange at a certain
scale $\lambda$, gets progressively shorter down the scales. In
this case the decay is governed by the slowest kinetics at
$\lambda_\mathrm{en}$, where the energy flux (per unit mass)
$\varepsilon$ is formed, while the faster kinetic processes at
shorter scales are able to instantly adjust to this flux
supporting the transfer of energy between neighboring scales
towards $\lambda_\mathrm{cutoff}$. Thus, the cascade is a
(quasi-)stationary regime in which the energy flux $\varepsilon$
is constant through the length scales, while the variation of
$\varepsilon$ in time happens on the longest time scale of the
system $\tau_{\rm coll} (\lambda_\mathrm{en})$.

\begin{figure}[htb]
\includegraphics[width = 0.75\columnwidth,keepaspectratio=true]{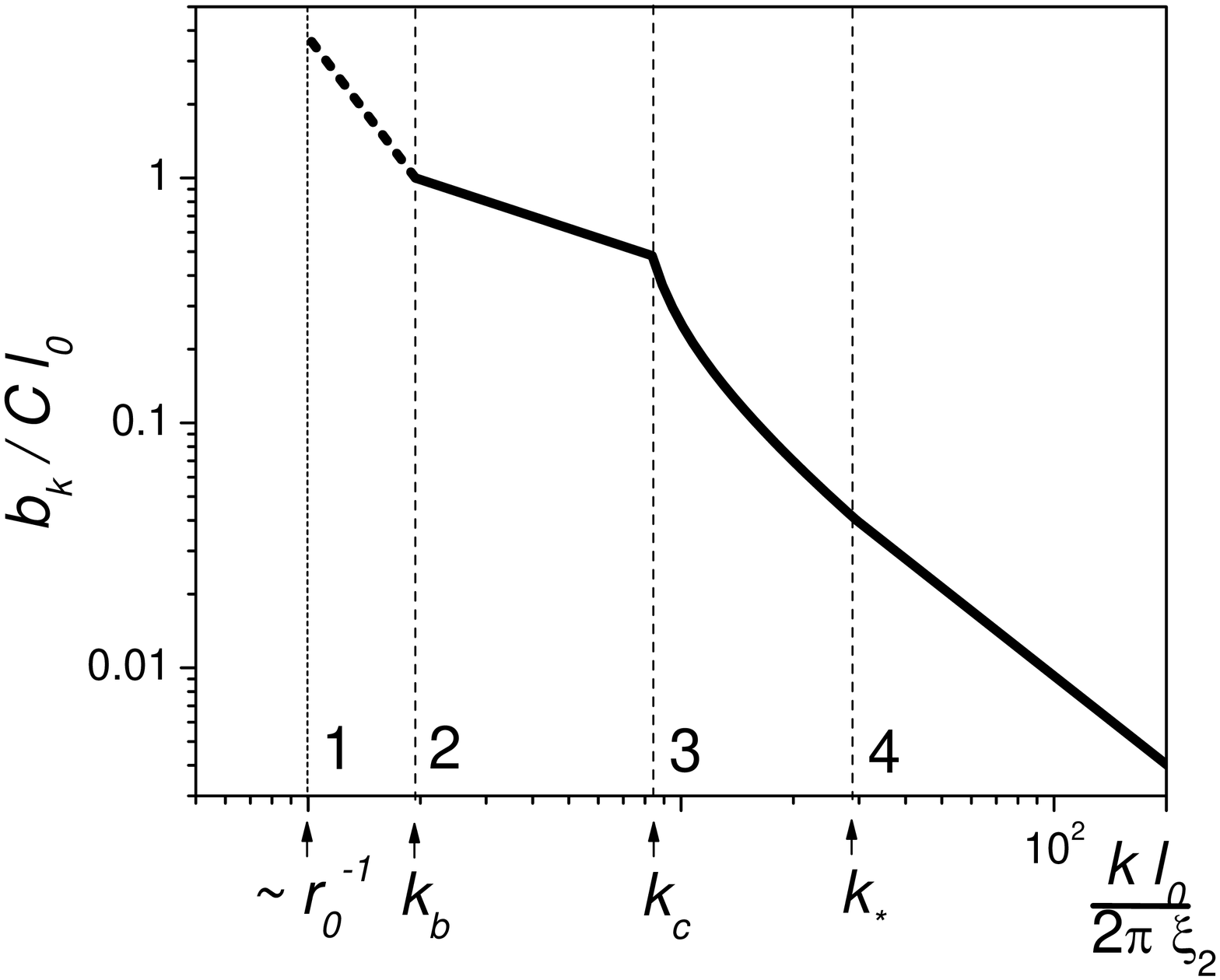}
\caption{Spectrum of Kelvin waves in the quantized regime as
predicted in Ref.~\cite{KS_07} and quantified here, apart from the
regime (1), by the fit to experimental data, Fig.~\ref{f2}. In
view of Eq.~(\ref{line_density}), the constants $C$ and $\xi_2$
can be found only in the combination $\xi_2 C \approx 0.049$. The
inertial range consists of a chain of cascades driven by different
mechanisms: (1) reconnections of vortex-line bundles, (2)
reconnections between nearest-neighbor vortex lines in a bundle,
(3) self-reconnections on single vortex lines, (4) non-linear
dynamics of single vortex lines without reconnections. The regimes
(3) and (4) are familiar in the context of non-structured vortex
tangle decay \cite{Sv_95, KS_04}. }
\label{f1}
\end{figure}

In Ref.~\cite{KS_07} we argued that at the effective zero
temperature the inertial range is highly nontrivial, since the
energy transport mechanisms change several times within it. We
first briefly outline the $T=0$ case (see \cite{KS_07} for
details), and then proceed to include the effect of mutual
friction due to the finite $T$. In this scenario, the key
dynamical parameter is
\begin{equation} \Lambda\,  =\,  \ln(l_0/a_0)\, \gg\, 1\; ,
\label{Lambda}
\end{equation}
where $a_0$ is the vortex core radius and $l_0$ is the typical
separation between the vortex lines determined by the Kolmogorov
energy flux $\varepsilon$ \cite{KS_07}. The parameter $\Lambda$
controls the competition between the local self-induced motion of
the vortex lines and the coupling between them. At the largest
length scales in the inertial range, $\lambda \gg r_0 \sim
\Lambda^{1/2} l_0$, the vortex lines are coupled in bundles that
move coherently mimicking the velocity profile of the
classical-fluid Kolmogorov turbulence. (The indistinguishability
of ST from its classical counterpart at large scales has been
attracting much attention since it was first predicted in
Ref.~\cite{Nore}.) At the scale $r_0$, the self-induced motion of
vortex lines becomes appreciable and the cascade enters the
\textit{quantized} regime, where each vortex line evolves
independently (apart from reconnections). Before the purely
non-linear KW cascade \cite{KS_04} can develop on separate vortex
lines at $k \gg k_* \sim \Lambda^{1/2}/l_0$ the cascade
experiences a complex transformation stage necessary to crossover
from spatially organized bundles at $k \ll r_0^{-1}$ and to
generate the KWs. In the crossover, which (given $\Lambda\sim 10$)
extends for about a decade in the wavenumber space, the
vortex-line reconnections \cite{Sv_95} become the main driving
mechanism of the energy transport in the cascade. Remarkably, the
crossover range is further split into three (sub-)regimes
distinguished by their specific types of reconnections: (1)
reconnections of vortex-line bundles ($r_0^{-1} \ll k \ll
k_\mathrm{b} \sim 1/\Lambda^{1/4}l_0$); (2) reconnections between
nearest-neighbor lines ($k_\mathrm{b} \ll k \ll k_\mathrm{c} \sim
\Lambda^{1/4}/l_0$); (3) self-reconnections on single lines
($k_\mathrm{c} \ll k \ll k_*$).
It is important here that each regime features a
distinct spectrum of KW amplitude $b_k$ summarized in
Fig.~\ref{f1}.

If the vortex lines were smooth, the vortex-line density $L$ would
be simply related to the interline separation as $L = l_0^{-2}$.
The increase of $L$ due to the creation of KWs on the vortex lines
is related to their spectrum by \cite{Sv_95}
\begin{equation}
\ln \left[L(\alpha)/L_0\right] \; = \;
\int_{\tilde{k}}^{k_\mathrm{cutoff}(\alpha)} (b_k k)^2 \; dk/k \;
. \label{line_density}
\end{equation}
Here $k_\mathrm{cutoff}\sim 1/\lambda_\mathrm{cutoff}$,
$\tilde{k}$ is the smallest wavenumber of the KW cascade (not to
be confused with the smallest wavenumber of the Kolmogorov
cascade) at which the concept of a definite cutoff scale is
meaningful, and $L_0$ is the ``background" line density
corresponding to $k_\mathrm{cutoff} \sim \tilde{k}$. There is an
ambiguity in the definition of $b_k$ associated with the choice of
the spectral width of the scale $k$, which is fixed in
Eq.~(\ref{line_density}) by setting the proportionality constant
between the l.h.s. and the r.h.s. to unity.

At $T=0$, the cascade is cut off by the radiation of sound (at
least in $^4$He) \cite{Vinen2000} at the length scale
$\lambda_\mathrm{cutoff} = \lambda_\mathrm{ph}$
\cite{KS_05_vortex_phonon}, \cite{KS_07}. As we show below,
changing the temperature one controls $\lambda_\mathrm{cutoff}(T)
> \lambda_\mathrm{ph}$ in Eq.~(\ref{line_density}), which allows one
to \textit{scan} the KW cascade observing qualitative changes in
$L(T)$ as $\lambda_\mathrm{cutoff}$ traverses different cascade
regimes. The existence of a well-defined cutoff is due to the fact
that the cascade is supported by rare kinetic events in the sense
that $\tau_{\rm coll}\equiv \tau_{\rm coll}(\varepsilon, k)$ is
much larger than the KW oscillation period, $\tau_{\rm per}\equiv
\tau_{\rm per}(k)$. The dissipative time $\tau_{\rm dis}\equiv
\tau_{\rm dis}(\alpha, k)\sim \tau_{\rm per} /\alpha$, as we show
below, is the typical time of the frictional decay of a KW at the
scale $k$. Thus, the cutoff condition is
\begin{equation}
\tau_{\rm dis}(\alpha, k)\, \sim \tau_{\rm coll}(\varepsilon, k) \; , \label{cutoff}
\end{equation}
which implies that the energy dissipation rate at a
given wavenumber scale becomes comparable to the energy being
transferred to higher wavenumber scales per unit time by the
cascade. It is this condition that defines the cutoff wavenumber
$k_{\rm cutoff}\equiv k_{\rm cutoff}(\varepsilon, \alpha)$. Decreasing
$T$ and thus $\alpha(T)$, one gradually increases $k_{\rm cutoff}$
thereby scanning the cascade. In view of
Eq.~(\ref{line_density}), this in principle allows one to extract
the KW spectrum.

At finite $T$, dissipative dynamics of a vortex line element is
described by the equation (omitting the third term in the r.h.s.,
which is irrelevant for dissipation) \cite{Donnelly,Cambridge}
\begin{equation}
\dot{\mathbf{s}}=\mathbf{v}(\mathbf{s})+\alpha\mathbf{s}'\times\left[\mathbf{v}_\mathrm{n}(\mathbf{s})-\mathbf{v}(\mathbf{s})\right]\;
. \label{BS+friction}
\end{equation}
Here $\mathbf{v}(\mathbf{r})$ is the superfluid velocity field,
$\mathbf{v}_\mathrm{n}(\mathbf{r})$ is the normal velocity field,
$\textbf{s}=\textbf{s}(\xi, t)$ is the time-evolving radius-vector
of the vortex line element parameterized by the arc length, the
dot and the prime denote differentiation with respect to time and
the arc length, respectively.

At $\alpha \sim 1$ the superfluid and normal components are
strongly coupled and the KWs are suppressed. In this
case, the cascade must cease \textit{before} it enters the quantized
regime, i.e.
$\lambda_\mathrm{cuttoff} \gtrsim r_0$. In this Letter, we do not
discuss the strongly coupled case, since it has no control
parameter for a theory. Consideration regarding energy spectra in
this regime can be found in Ref.~\cite{En_spectra}.

If, however, the mutual friction is small, then $\alpha^{-1} \gg
1$ gives the characteristic number of KW oscillations required for
the wave to decay. Indeed, to the first approximation in
$1/\Lambda$, $\mathbf{v}(\mathbf{s})$ in Eq.~(\ref{BS+friction})
is given by the local induction approximation (LIA)
\cite{Donnelly},
\begin{equation}
\mathbf{v}(\textbf{s}) \approx  \Lambda_{R} \, {\kappa \over 4
\pi}  \, \textbf{s}' \times \textbf{s}'', \;\;\;\;\;  \Lambda_{R} =
\ln (R/a_0)\; , \label{LIA}
\end{equation}
where $\kappa$ is the circulation quantum and $R$ is the typical
curvature radius. Taking into account that $\Lambda_R$ is a very
weak function of $R$, we shall treat it as a constant of the typical
value $\Lambda_{R} \sim \Lambda$. As long as $\alpha \ll 1$, the
disturbance of the normal component caused by vortex line motion
can be neglected in Eq.~(\ref{BS+friction}), so that
$\mathbf{v}_\mathrm{n}(\mathbf{s})$ is irrelevant for KW
dissipation. In this case, Eqs.~(\ref{BS+friction}), (\ref{LIA})
give the rate at which the amplitude $b_k$ of a KW with
wavenumber $k$ decays due to the mutual friction:
\begin{equation}
\dot{b}_k \, \sim \, - \, \alpha \, \omega_k \, b_k\; ,
\label{amplitude_decay}
\end{equation}
while the KW dispersion is $\omega_k=(\kappa/4\pi)\Lambda
k^2$. Here and below we omit factors of order unity, which are
subject to the definition of the spectral width of the wave. These
factors can not be found within our theory, but can be extracted
from experimental data (as we do it below) and from numerical
simulations. Since the energy per unit line length associated with
the wave is $E_k \sim \kappa \rho \omega_k b_k^2$, the power
dissipated (per unit line length) at the scale $\sim k^{-1}$ is
given by
\begin{equation}
\Pi(k) \, \sim \, \alpha \, \kappa \, \rho \, \omega_k^2 \, b_k^2, \label{power}
\end{equation}
where $\rho$ is the fluid density. In the following, we analyze ST
as $\alpha(T)$ scans through the regimes shown in Fig.~\ref{f1}.

\textit{Regime (1).} As it was already mentioned, at this stage
the vortex lines are organized in bundles; the amplitudes of waves
on these lines are given by \cite{KS_07} $b_k \sim r_0^{-1}
k^{-2}$. One can estimate the total power lost due to the friction
of these bundles at the wavenumber scale $r_0^{-1} \ll k \ll
k_\mathrm{b}$ per unit mass of the fluid as
\begin{equation}
\varepsilon_\mathrm{dis}(k)\, =\, \frac{1}{\rho b_k^2} \;
\frac{b_k^2}{l_0^2} \; \Pi(k) \; \sim \; \alpha \, \kappa^3 \,
\Lambda \, / \, l_0^4 \; .\label{reg1}
\end{equation}
Here, the first factor in the r.h.s. is associated with the
correlation volume at this scale $\sim b_k^2 k^{-1}$ \cite{KS_07}
and the second one stands for the number of vortex lines in the
volume. Note that the dissipated power is constant at all the
length scales within this regime and, since according to
Ref.~\cite{KS_07} the interline separation is related to the
Kolmogorov flux by $\varepsilon \sim \Lambda \kappa^3/l_0^4$, it
is simply given by $\alpha \varepsilon$. Thus, when $\alpha \ll 1$
the kinetic channel in the whole regime (1) becomes efficient and
the cascade reaches the scale $\lambda_\mathrm{b}$, where the
notion of bundles becomes meaningless. That is the regime (1), as
opposed to the regimes (2)-(4), is \textit{not} actually scanned
by $\alpha$. Basically, the inertial range of the regime (1)
develops as a whole while $\alpha$ evolves from the values of
order unity to the values much smaller than unity.

\textit{Regime (2).} In the range $k_\mathrm{b} \ll k \ll k_\mathrm{c}$ ,
the spectrum of KWs is given by \cite{KS_07} $b_k \sim
l_0 (k_\mathrm{b}/k)^{1/2}$, which for the dissipated power yields
\begin{equation}
\varepsilon_\mathrm{dis}(k)  \sim \frac{1}{\rho l_0^2} \, \Pi(k)
\sim \alpha \kappa^3 \Lambda^{7/4}k^{3}/l_0\; . \label{reg2}
\end{equation}
The condition $\varepsilon_\mathrm{dis}(k) \sim \varepsilon$ gives
the cutoff wavenumber
\begin{equation}
k_\mathrm{cutoff} \; = \; \xi_2 \, \Lambda^{-1/4} \alpha^{-1/3} \,
(2\pi/l_0)\; , \;\;\; k_\mathrm{b} \ll k_\mathrm{cutoff} \ll k_\mathrm{c}\;
, \label{cutoff_reg2}
\end{equation}
where $\xi_2$ is some constant of order unity.  Then, from
Eq.~(\ref{line_density}) we obtain
\begin{equation}
\ln\frac{L(k_\mathrm{cutoff})}{L_0} = C^2 (k_\mathrm{b} l_0)^2
[k_\mathrm{cutoff}/k_\mathrm{b} - 1]\; .\label{L_reg2}
\end{equation}
Here, we set $b_k=C \, l_0 (k_\mathrm{b}/k)^{1/2}$, where C is a constant of order unity. The overall
magnitude of $b_k$ in the other regimes follows then from the
continuity.

\textit{Regime (3).} In this regime, supported by
self-reconnections, the spectrum is given by $b_k \sim k^{-1}$ (up
to a logarithmic pre-factor). The corresponding energy balance
condition yields the cutoff scale:
\begin{equation}
k_\mathrm{cutoff} \; = \; \xi_3 \,(\Lambda \alpha)^{-1/2}\,(2
\pi/l_0), \;\;\; k_\mathrm{c} \ll k_\mathrm{cutoff} \ll k_* \; .
\label{cutoff_reg3}
\end{equation}
With the logarithmic pre-factor taken into account \cite{Sv_95},
the spectrum in this regime reads $b_k=C
[1+c_3^2\ln(k/k_\mathrm{c})]^{-1/2}( \sqrt{ k_\mathrm{c} k_\mathrm{b} }/k)l_0$,
where $c_3$ is a constant of order unity. Then, the relative
increase of vortex line density through this regime is given by
\begin{equation}
\frac{L(k_\mathrm{cutoff})}{L(k_\mathrm{c})} \; =  \; \left[ \, 1 \; + \;
c_3^2 \; \ln\frac{k_\mathrm{cutoff}}{k_\mathrm{c}}\, \right]^{\nu}\; ,
\label{L_reg3}
\end{equation}
where $\nu = C^2 k_\mathrm{c} k_\mathrm{b} l_0^2/c_3^2$.

\textit{Regime (4).} Since the spectrum of the purely nonlinear
regime, $b_k \sim k^{-6/5}$, is steeper than the marginal $b_k
\sim k^{-1}$ meaning that the integral in Eq.~(\ref{line_density})
builds up at the lower limit, as soon as $k_\mathrm{cutoff}
\gtrsim k_*$ the line density $L(k_\mathrm{cutoff})$ starts to
saturate and becomes independent of $k_\mathrm{cutoff}$ at
$k_\mathrm{cutoff} \gg k_*$. The energy balance gives the
dependence $k_\mathrm{cutoff}$ on $\alpha$ in the form
\begin{equation}
k_\mathrm{cutoff} = \xi_4 \, \Lambda^{-3/4} \, \alpha^{-5/8} \,
(2\pi/l_0), \;\;\;  k_\mathrm{cutoff} \gg k_*  \; ,
\label{cutoff_reg4}
\end{equation}
where $\xi_4$ is an unknown constant. The coefficient in the
KW spectrum is fixed by continuity with the previous
regime, yielding $b_k =C [1+c_3^2 \ln(k_*/k_\mathrm{c})]^{-1/2} \sqrt{k_\mathrm{c}
k_\mathrm{b} } \, k_*^{1/5} k^{-6/5} l_0$.
Eq.~(\ref{line_density}) thus yields
\begin{equation}
\frac{L(k_\mathrm{cutoff})}{L(k_*)} \approx 1+  \frac{(5C^2\,/2) k_\mathrm{c} k_\mathrm{b}l_0^2}{1+c_3^2
\ln(k_*/k_\mathrm{c})} \left[ 1-
\left(\frac{k_*}{k_\mathrm{cutoff}}\right)^{2/5}\right]\; .
\label{L_reg4}
\end{equation}

The continuity of $k_\mathrm{cutoff}$ leads to the following
constraints on the  free coefficients in Eqs.~(\ref{cutoff_reg2}),
(\ref{cutoff_reg3}), (\ref{cutoff_reg4}), $\xi_3\;=\;\xi_2\;
\,\Lambda^{1/4} \, \alpha_\mathrm{c}^{1/6}$, $\xi_4\;=\;\xi_2\;
\Lambda^{1/2}\,\alpha_\mathrm{c}^{1/6}\alpha_*^{1/8}$, where
$\alpha_\mathrm{c} \sim \Lambda^{-3/2}$ is the value of the
friction coefficient at which the cascade is cut off at the
crossover between the regimes (2) and (3) and $\alpha_* \sim
1/\Lambda^2$ corresponds to the one between (3) and (4).
Introducing $\alpha_b \lesssim 1$, corresponding to the crossover
from the regime (1) to (2), we rewrite Eqs.~(\ref{L_reg2}),
(\ref{L_reg3}), (\ref{L_reg4}) in terms of $\alpha$:
\begin{gather}
\ln \frac{L(\alpha)}{L_0} = A_2 \left[ (\alpha_\mathrm{b}/\alpha)^{1/3} -1 \right], \;\;\; \alpha_\mathrm{c} \ll \alpha \ll \alpha_b, \notag \\
\frac{L(\alpha)}{L(\alpha_\mathrm{c})}=\left[ \, 1 \; + \; (c_3^2/2) \; \ln\frac{\alpha_\mathrm{c}}{\alpha}\, \right]^{\nu}, \; \; \; \alpha_* \ll \alpha \ll \alpha_\mathrm{c}, \notag \\
\frac{L(\alpha)}{L(\alpha_*)} \approx 1+ A_4 \left[ 1-
\left(\frac{\alpha}{\alpha_*}\right)^{1/4}\right], \notag
\;\;\;\;\; \alpha \ll \alpha_*, \nonumber
\end{gather}
where $A_2= (2 \pi)^2 \xi_2^2 C^2/\Lambda^{1/2}
\alpha_\mathrm{b}^{2/3}$, $\nu = (2 \pi)^2 \xi_2^2
C^2/\Lambda^{1/2} (\alpha_\mathrm{c} \alpha_\mathrm{b})^{1/3}
c_3^2$, and $A_4=5(2 \pi)^2 \xi_2^2 C^2/2 \Lambda^{1/2}
[1+(c_3^2/2) \ln(\alpha_\mathrm{c}/\alpha_*)] (\alpha_\mathrm{c}
\alpha_\mathrm{b})^{1/3}$.

\begin{figure}[htb]
\includegraphics[width = 0.85\columnwidth,keepaspectratio=true]{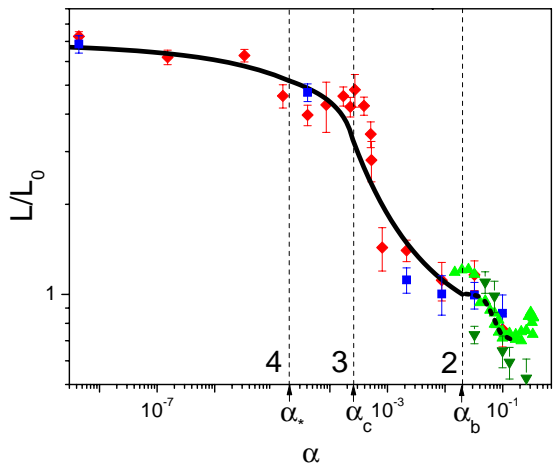}
\caption{(Color online.) Fit of the experimental data (squares and
diamonds) adapted from Ref.~\cite{Walmsley} ($l_0 \approx
5\times10^{-3} cm$ giving $\Lambda \approx 13$). The high
temperature measurements of Refs.~\cite{Oregon}, \cite{Ladik} are
represented by triangles and inverted triangles respectively.  The
form of $\alpha(T)$ is taken according to
Ref.~\cite{Samuels_Donnelly} at $T \gtrsim 0.5K$ (roton
scattering) and according to Ref.~\cite{Iordanskii} at $T \lesssim
0.5K$ (phonon scattering). The fitting parameters are
$A_2\approx0.36$, $A_4\approx0.33$, $c_3\approx2.9$,
$\nu\approx0.18$ with $\alpha_\mathrm{b} \approx 2 \times
10^{-2}$, $\alpha_\mathrm{c} \approx 2.5 \times 10^{-4}$,
$\alpha_* \sim 2\times10^{-5}$.}
\label{f2}
\end{figure}

In conclusion, we discuss the qualitative form of $L(\alpha)$ in
connection with the alternative scenario of the crossover from the
Kolmogorov to the KW cascade proposed by L'vov \textit{et al.}
\cite{Lvov}. In Ref.~\cite{KS_07} we  argued that this scenario
mistakenly leaves out the reconnection-driven regimes. In this
scenario the classical regime extends down to the scale of
interline separation $l_0$, where it is \textit{immediately}
followed by the purely non-linear KW cascade [regime (4) in our
notation]. Since the kinetics of the regime (4) are too weak to
support the flux $\varepsilon$ at this scale, the vorticity was
proposed in \cite{Lvov} to build up in the classical regime at the
scales adjacent to $l_0$---the so-called bottleneck effect.
Although this picture would also naturally predict a significant
increase of $L$, the peculiar behavior of $L(\alpha)$ observed in
\cite{Walmsley} essentially rules out the scenario of
Ref.~\cite{Lvov}. The reason is that the dramatic rise of $L$
happens only at $\alpha$ as low as $10^{-3}$, while the bottleneck
accumulation of classical vorticity must manifest itself as soon
as KWs become an unavoidable relaxational channel, i.e. already at
$\alpha \lesssim 1$.

Instead, $L(\alpha)$ exhibits the qualitative form peculiar to our
scenario of Ref.~\cite{KS_07}. As $\alpha$ decreases from $\alpha
\sim 1$ to the values significantly smaller than unity, the line
density $L$ increases only by some factor close to unity ($\sim
1.5$ in the experiment), which reflects the formation of the
regime (1) driven by the reconnections of vortex bundles. During
the crossover from the region (1) to (2), the increase of $L$ is
minimal---a \textit{shoulder} in the curve $L(\alpha)$ arises, in
contrast to a large increase of $L$ expected in the bottleneck
picture.  It is only well inside the region (2) that the increase
of $L$ becomes progressively pronounced, and, at the crossover to
the region (3), the function $L(\alpha)$ achieves its maximal
slope, determined by the fractalization of the vortex lines
necessary to support the cascade within the interval (3). As the
cutoff moves along the interval (3) towards higher wavenumbers,
the slope of $L(\alpha)$ becomes less steep due to the decrease of
the characteristic amplitude of KW turbulence. When the cutoff
passes the crossover to the regime (4), the curve $L(\alpha)$
gradually levels.
Fitting the experimental data fixes the values of all the
dimensionless parameters, thus revealing the {\it quantitative}
form of the KW spectrum, Fig.~\ref{f1}. Note, however, that in
view of the fact that the experimental $\Lambda \sim 10$ is not so
large we would expect certain systematic deviations, but a
significant scatter of the experimental data does not allow us to
assess them.

\end{document}